\documentclass{elsart}
 \usepackage{amssymb}
 \usepackage{epsfig}
    \newcommand{\ba}{\begin{eqnarray}}
    \newcommand{\ea}{\end{eqnarray}}
    \newcommand{\be}{\begin{equation}}
    \newcommand{\ee}{\end{equation}}

    \newcommand{\AmS}{{\protect\the\textfont2
  A\kern-.1667em\lower.5ex\hbox{M}\kern-.125emS}}

\begin{document}
\runauthor{Junhua and Chuan}
\begin{frontmatter}

\title{Static Quark Potential and the Renormalized Anisotropy
on Tadpole Improved Anisotropic Lattices\thanksref{fund}}
\author[Beida]{Wei Liu},
\author[IHEP]{Ying Chen},
\author[Beida]{Ming Gong},
\author[Beida]{Xin Li},
\author[Beida]{Chuan Liu},
\author[Beida]{and Guozhan Meng}
\address[Beida]{School of Physics, Peking University\\
          Beijing, 100871, P.~R.~China}
\address[IHEP]{Institute of High Energy Physics\\
                Academia Sinica, P.~O.~Box 918\\
                Beijing, 100039, P.~R.~China}
\thanks[fund]{This work is supported by the Key Project of National Natural
Science Foundation of China (NSFC) under grant No. 10235040, No.
10421503, and supported by the Trans-century fund and the Key
Grant Project of Chinese Ministry of Education (No. 305001).}

 \begin{abstract}
 Static quark potential is studied using a tadpole
 improved gauge lattice action. The scale is set using
 the potential for a wide range of bare parameters.
 The renormalized anisotropy of the lattice is also
 measured.
 \end{abstract}
 \end{frontmatter}


\section{Introduction}

 It has become clear that anisotropic lattices and improved
 lattice actions greatly facilitate lattice
 QCD calculations involving heavy physical objects like the glueballs,
 one meson states with non-zero three momenta and
 multi-meson states with or without three momenta.
 It is also a good workplace for the study of hadrons
 with heavy quarks. In many of the lattice calculations,
 it is useful to use finite size techniques in which
 lattices with fixed physical volumes are simulated.
 For the purpose of these calculations, one needs to
 know the correspondence between the lattice spacing
 in physical units and the bare gauge coupling $\beta$.
 It is also important to check the renormalization effects
 of the anisotropy parameter for these actions.
 In this work we present our numerical
 studies on these issues using the tadpole improved
 gluonic action on asymmetric lattices:
 \ba
 \label{eq:pure_gauge}
 S=&-&\beta\sum_{i>j} \left[
  {5\over 9}{TrP_{ij} \over \xi_0 u^4_s}
 -{1\over 36}{TrR_{ij} \over \xi_0 u^6_s}
 -{1\over 36}{TrR_{ji} \over \xi_0 u^6_s} \right] \nonumber \\
 &-&\beta\sum_{i} \left[ {4\over 9}{\xi_0 TrP_{0i} \over  u^2_s}
 -{1\over 36}{\xi_0 TrR_{i0} \over u^4_s} \right] \;\;£¬
 \ea
 where $P_{ij}$ is the usual plaquette variable and
 $R_{ij}$ is the $2\times 1$ Wilson loop on the lattice.
 The parameter $u_s$, which we take to be the forth root
 of the average spatial plaquette value, incorporates the
 usual tadpole improvement~\cite{lepage93:tadpole} and
 $\xi_0$ designates the bare
 aspect ratio of the anisotropic lattice.
 With the tadpole improvement in place, the bare anisotropy
 parameter $\xi_0$ suffers only small renormalization.
 Therefore, the physical, or renormalized anisotropy parameter
 $\xi$, which can be determined by comparing physical quantities
 which depend on distances in both the spatial and the
 temporal directions, will be quite close to its bare value $\xi_0$.
 The effect of this renormalization will be studied in
 this paper for various values of $\beta$ using the so-called
 side-way potential method.
 Using the pure gauge action~(\ref{eq:pure_gauge}),
 glueball and light hadron spectrum has been studied
 within the quenched approximation \cite{colin97,colin99,%
 chuan01:gluea,chuan01:glueb,chuan01:canton1,chuan01:canton2,chuan01:india}.

 In this paper, we will study the static quark anti-quark
 potential using the pure gauge action given in
 Eq.~(\ref{eq:pure_gauge}). The static quark anti-quark potential
 is obtained by the measurement of the Wilson loops on the lattice.
 Then, we set the scale in
 physical unit using the Sommer scale $r_0$ which is defined via
 the static quark potential. This establishes the relation
 between the lattice spacing and the bare gauge coupling.
 For all the $\beta$ values we studied, we also investigate
 the renormalized anisotropy parameter $\xi$
 using the side-way potential method.
 Similar studies have been carried out
 before~\cite{alford01:aniso_xi,sakai04:aniso_xi}. Our study
 covers more parameter space which is used in anisotropic lattice
 simulations. The renormalized anisotropy is also
 calculated within perturbation theory to
 one-loop~\cite{drummond02:aniso_xi}.

 This paper is organized in the following manner. In Section 2,
 parameters of our simulations are given and the details of
 the Wilson loop measurements are presented. From these data,
 we extract the static quark potential and the lattice spacing
 is determined in terms of the physical scale $r_0$.
 The results for all $\beta$ are then interpolated using
 quadratic polynomials. These interpolation then offers a
 rather precise correspondence between $\beta$ and $r_0/a_s$ over the
 whole range of $\beta$ that have been simulated.
 In section 3, we discuss the side-way potential
 measurements which yield the
 renormalized anisotropy for all our simulation points.
 In Section 4, we will conclude with some general remarks.

 \section{Simulation Results for Static Potential}

 In this section, we present our numerical results for
 the study of the static quark anti-quark potential
 for various gauge coupling $\beta$.
 Our values of $\beta$ range from $1.9$ to $2.8$ which
 roughly corresponds the lattice spacing $a_s$ in
 the range of $0.35$fm to $0.12$fm. Details of the
 simulation parameters are summarized in Table 1.

 Since we are interested in the spatial lattice spacing
 in physical unit, which is sensitive to the ultra-violet
 physics, we therefore only use small lattices of size $8^3\times 40$.
 The lattices are generated using the conventional
 Cabbibo-Mariani pseudo-heatbath algorithm with
 over-relaxation. In a compound sweep, we perform one pseudo-heatbath
 sweep with three over-relaxation sweeps. For each $\beta$, over a thousand of
 gauge field configurations are generated, with each
 configuration separated by 3 compound sweeps mentioned above.

 Using these gauge field configurations, Wilson loops
 are constructed and measured which yield static quark anti-quark potential.
 It is known that, Wilson loops, especially large loops are extremely noisy
 objects. We thus utilize the conventional smearing techniques for the spatial
 gauge links. In such a process, each spatial gauge link is replaced by its
 original value plus a linear combination of its nearest staples with
 a coefficient $\lambda_s$. Finally, each spatial gauge link is
 projected back into group $SU(3)$. Two different sets of smearing parameters
 are chosen for small spatial distance $R$ and large $R$.
 Typically  the smearing parameter
 in our simulations lies in the range: $\lambda_s=0.1-0.4$.
 Smearing can be done iteratively.
 We usually perform 4 or 6 sweeps of smearing before
 our construction of the Wilson loops.
 Another useful technique is to use thermally averaged temporal
 links. This also greatly reduces the statistical errors.

 The Wilson loops\ $W(R,t)$\ at a fixed spatial distance $R$ and
 large temporal separation $t$ is related to the static
 quark anti-quark potential $V(R)$ by:
 \be
 \label{eq:wilson_loop}
 W(R,t) \simeq e^{-tV(R)}= e^{-\hat{t}\hat{V}(R)}\;,\;\;
 t\rightarrow \infty\;.
 \ee
 Note that here we have used dimensionless temporal
 separation $\hat{t}=t/a_t$ which assumes integral values.
 What is really measured is the dimensionless potential:
 $\hat{V}=a_tV(R)$. Within the range that we can reach in our simulation,
 we find that the potential can be well-represented by a linear term plus
 a Coulomb term. Therefore, at various values of $R$,
 we fit the potential $V(R)$ using the following form:
 \be
 V(R)=V_0+{\alpha \over R} +\sigma R\;,
 \ee
 Or, in terms of the dimensionless potential:
 \be
 \label{eq:potential_dimensionless}
  \hat{V}(\hat{R})=\hat{V}_0+{\hat{\alpha} \over \hat{R}}
  + \hat{\sigma}\hat{R}\;,
 \ee
 with $\hat{V}_0=a_tV_0$, $\hat{\alpha}=\alpha a_t/a_s$,
 and $\hat{\sigma}=\sigma a_ta_s$.

 At this stage, we tried two ways of extracting the
 parameters $\hat{\alpha}$ and $\hat{\sigma}$.
 In the first method, one first checks for plateau behaviors
 of ratios of Wilson loops between two neighboring $\hat{t}$
 in the large temporal region. According to Eq.~(\ref{eq:wilson_loop}),
 the height of these plateaus
 will give us the estimate for the dimensionless potential
 $\hat{V}(\hat{R})$ for all values of $\hat{R}$. Then, in the
 second step, one fits the resulting $\hat{V}(\hat{R})$ versus
 $\hat{R}$ using Eq.~(\ref{eq:potential_dimensionless}). The
 result of this fit then gives the optimal values for $\alpha$
 and $\sigma$. In the second method, one directly performs a
 correlated fit for the on-axis Wilson loops using the form:
 \be
 W(R,t)=Z(\hat{R})
 \exp\left[-\hat{t}\left(\hat{V}_0+{\hat{\alpha}\over \hat{R}}
 +\hat{\sigma} \hat{R}\right)\right]\;,
 \ee
 Note that here all the $Z(\hat{R})$ parameters plus the
 three parameters: $\hat{V}_0$, $\hat{\alpha}$ and $\hat{\sigma}$
 enters the fitting process.

 \begin{figure}[htb]
 \begin{center}
 \includegraphics[width=12.0cm,angle=0]{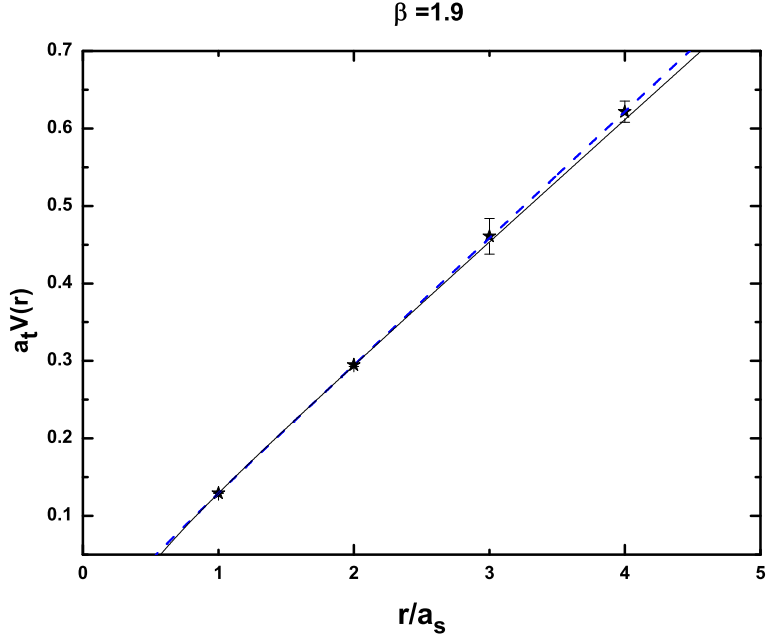}
 \end{center}
 \caption{
 The dimensionless potential $\hat{V}(\hat{R})$ is fitted versus $\hat{R}$
 using Eq.~(\ref{eq:potential_dimensionless}) for $\beta=1.9$. The dashed
 curve is the fit using the first method while the solid curve represents
 a direct fitting to the Wilson loops.}
 \label{fig:potential_fit19}
 \end{figure}
 \begin{figure}[htb]
 \begin{center}
 \includegraphics[width=12.0cm,angle=0]{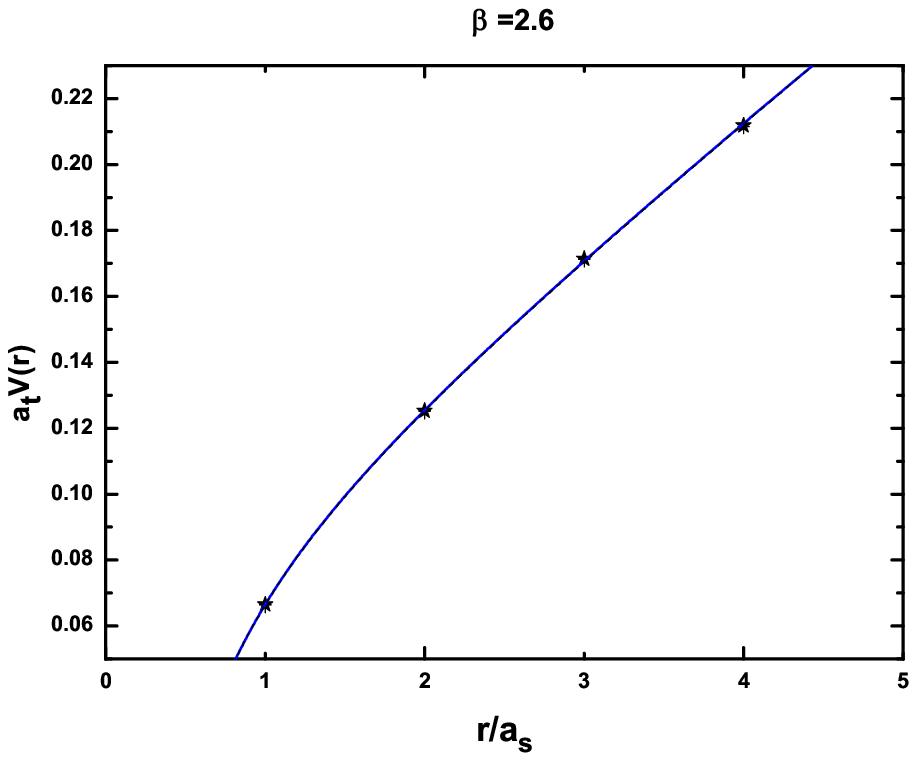}
 \end{center}
 \caption{Same as Fig.~\ref{fig:potential_fit19}, but for $\beta=2.6$. }
 \label{fig:potential_fit26}
 \end{figure}
 In Fig.~\ref{fig:potential_fit19} and Fig.~\ref{fig:potential_fit26},
 we show the fit of the static potential
 for two different values of $\beta$.
 The dashed curves are the fits for the potential using the
 first method. For comparison purposes, potentials using
 the fitted parameters obtained with the second method are
 also plotted as the solid curves.
 The difference for the two methods is indistinguishable for
 large values of $\beta$. In any case, we find that the two
 methods always yield compatible results for the fitted parameters.
 We list in Table~\ref{tab:parameters} the fitted
 parameters $\hat{\alpha}$ and $\hat{\sigma}$ for
 all $\beta$ values together with the $\chi^2$ per degree
 of freedom of the fit.

 \begin{table}[htb]
 \caption{Simulation parameters used in this work.
 All lattices are of size $8^3\times 40$ and $\xi_0=5$.
 \label{tab:parameters}}
 \begin{center}
 \begin{tabular}{|c||c|c|c|c|c|c|c|c|c|c|}
 \hline
 $\beta$ &
 $1.9$ & $2.0$ & $2.1$ & $2.2$ & $2.3$ \\
 \hline
 \hline
 $u^4_s$ &
 $0.328$ & $0.345$ & $0.361$ & $0.377$ & $0.394$ \\
 \hline
 $\hat{\alpha}$ &
 $-0.017(2)$ & $-0.021(2)$ & $-0.027(2)$ & $-0.030(1)$ & $0.040(1)$ \\
 \hline
 $\hat{\sigma}$ &
 $0.157(1)$ & $0.1335(9)$ & $0.1123(7)$ & $0.0937(6)$ & $0.0733(6)$ \\
 \hline
 $\chi^2/d.o.f$ &
 $1.45$ & $1.04$ & $1.58$ & $0.89$ & $0.48$ \\
 \hline
 $r_0/a_s$ &
 $1.415(12)$ & $1.522(12)$ & $1.644(16)$ & $1.787(15)$ & $1.991(18)$ \\
 \hline
 \hline
 $A$ &
 $14(2)$ & $14(2)$ & $14(2)$ & $17(2)$ & $20(2)$ \\
 \hline
 $B$ &
 $-44(7)$ & $-44(7)$ & $-44(7)$ & $-58(9)$ & $-71(11)$ \\
 \hline
 $C$ &
 $39(8)$ & $39(8)$ & $39(8)$ & $54(9)$ & $69(12)$ \\
 \hline
 \hline
 $\beta$ &
 $2.4$ & $2.5$ & $2.6$ & $2.7$ & $2.8$ \\
 \hline
 \hline
 $u^4_s$ &
 $0.409$ & $0.424$ & $0.437$ & $0.451$ & $0.463$ \\
 \hline
 $\hat{\alpha}$ &
 $-0.0389(9)$ & $-0.0407(6)$ & $-0.0410(3)$ & $-0.0599(5)$ & $-0.0582(5)$ \\
 \hline
 $\hat{\sigma}$ &
 $0.0606(4)$ & $0.0482(3)$ & $0.0384(2)$ & $0.0282(2)$ & $0.0225(2)$ \\
 \hline
 $\chi^2/d.o.f$ &
 $0.90$ & $1.94$ & $2.81$ & $1.88$ & $0.83$ \\
 \hline
 $r_0/a_s$ &
 $2.191(17)$ & $2.449(17)$ & $2.743(11)$ & $3.093(20)$ & $3.476(23)$ \\
 \hline
 \hline
 $A$ &
 $23(2)$ & $31(3)$ & $34(3)$ & $34(3)$ & $34(3)$ \\
 \hline
 $B$ &
 $-88(11)$ & $-125(15)$ & $-142(16)$ & $-142(16)$ & $-142(16)$ \\
 \hline
 $C$ &
 $88(13)$ & $135(19)$ & $155(20)$ & $155(20)$ & $155(20)$ \\
 \hline
 \end{tabular}
 \end{center}
 \end{table}

 To set the scales of a lattice in physical unit,
 we use the so-called Sommer scale $r_0$ defined as:
 \be
 R^2\left.{dV(R)\over dR}\right|_{R=r_0}=1.65\;.
 \ee
 This implies that:
 \be
 {r_0\over a_s}=\sqrt{{\alpha+1.65\over \sigma a^2_s}}
 =\sqrt{{\hat{\alpha}+1.65/\xi\over \hat{\sigma}}}\;.
 \ee
 where $\xi=a_s/a_t$ is the (renormalized) anisotropy of the lattice.
 As we will verify, the renormalization effects for $\xi_0$ are
 quite small, therefore we can replace $\xi$ in the above
 expression by $\xi_0$ in most cases.
 Also in Table~\ref{tab:parameters}, we list the values of $(r_0/a_s)$ for
 various $\beta$ values.

 \begin{figure}[htb]
 \begin{center}
 \includegraphics[width=12.0cm,angle=0]{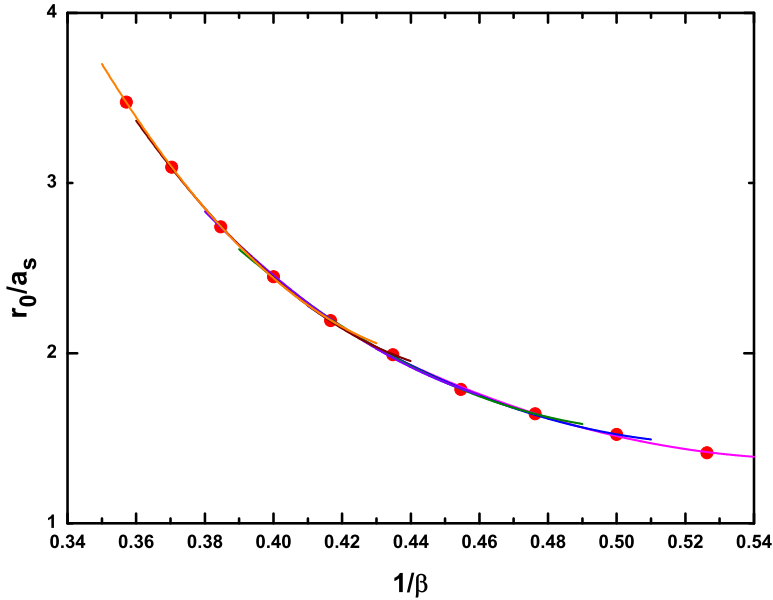}
 \end{center}
 \caption{Interpolations in the whole range of $\beta$ that we simulated.
 Around each simulated point, we perform a five point quadratic interpolation
 around that point. These interpolations are shown as colored lines. It is seen
 that these lines interconnect each other well, showing that consistency of
 the interpolation. }
 \label{fig:interpolation}
 \end{figure}
 In order to obtain a one-to-one correspondence between
 $\beta$ and $r_0/a_s$ within the whole range we studied,
 we have also attempted to make interpolations within
 the whole range of $\beta$ in our simulation.
 We use a five point quadratic fit in $1/\beta$ around each point to
 interpolate points between the neighboring points of $\beta$ that
 have been simulated:
 \be
 {r_0\over a_s}=A+{B\over \beta}+ {C\over \beta^2}\;.
 \ee
 The situation is shown in Fig.~\ref{fig:interpolation}.
 The data points are the resulting values of $r_0/a_s$ obtained
 from the static quark potential at various values of $\beta$.
 The colored lines are the quadratic fitting curves around
 each simulated point. We see that these quadratic interpolations
 are smoothly connected with one another. Within each interval,
 different fitting curves agree within errors.
 For reference, we tabulated the fitted parameters $A$, $B$, $C$ for
 each interpolation in Table~\ref{tab:parameters}.
 With these parameters, one can find the correspondence between
 $\beta$ and the physical lattice spacing within the whole
 range of interest.

 \section{Determination of renormalized anisotropy}

 We now move to the measurements of the anisotropy
 parameter $\xi=a_s/a_t$. This is a physical quantity
 which can be measured by various methods. In this
 work, we use the side-way potential method.
 The strategy for this method is briefly described below.

 On anisotropic lattices, one can measure both the spatial-temporal
 Wilson loops and the spatial-spatial Wilson loops.
 Both of these objects contain information about
 the static quark anti-quark potential as a function
 of the quark distances. Since there is one unique physical potential,
 by comparing the results from these two measurements, one infers the
 information about physical anisotropy.

 To be more precise, we assume that the spatial lattice
 axis are called $x$, $y$ and $z$, the temporal axis is called $t$.
 We can measure a spatial-temporal Wilson loop with the spatial
 separation $\mathbf{R}=R\hat{z}$ lies in the $z$ direction, with
 $\hat{z}$ being a unit vector in the $z$ direction. Since we
 are on a lattice, it is obvious that $R=\hat{R}a_s$, with
 $\hat{R}$ being positive integers.
 If we measure the spatial-temporal Wilson loop $W(\hat{R}a_s,\hat{t}a_t)$
 and we investigate the large $\hat{R}$ behavior of the loop, we have:
 \be
 W(Ra_s,ta_t)\propto e^{-\hat{R}a_sV_t(\hat{t}a_t)}\;, \;
 R\gg 1\;.
 \ee
 Therefore, at large values of $\hat{R}$, we obtain the
 combination $a_sV_t(\hat{t}a_t)$ for all values of $\hat{t}$.
 Here we use the notation $V_t$ to indicate that it is
 a potential obtained from a spatial-temporal Wilson loop
 measurement.

 On the other hand, we can also measure the
 spatial-spatial Wilson loop $W(\hat{R}a_s, \hat{R}'a_s)$
 with $\hat{R}'a_s$ indicating the distance
 between the quark and anti-quark separation in the $xy$ plane.
 We thus have:
 \be
 W(\hat{R}a_s,\hat{R}'a_t)\propto
 e^{-\hat{R}a_sV_s(\hat{R}'a_s)}\;, \;
 R\gg 1\;.
 \ee
 Again, for large enough $\hat{R}$, we obtain the
 potential $a_sV_s(\hat{R}'a_s)$. Here we use $V_s$ to
 indicate that it is the potential from a spatial-spatial
 Wilson loop measurement.
 Note, however, $V_t$ and $V_s$ are in fact the same physical
 quantity. That is to say, if they evaluated at the same
 physical distance, they ought to be identical.
 Therefore, by measuring the spatial-spatial and spatial-temporal
 Wilson loops, we obtain two versions of the same potential.
 The renormalized anisotropy $\xi$ has to be such that:
 \be
 \label{eq:VT}
 V(\hat{R}' a_s)=V({\hat{t}\over \xi} a_s)\;,
 \ee
 holds for all values of $\hat{R}'$.
 \begin{figure}[htb]
 \begin{center}
 \includegraphics[width=12.0cm,angle=0]{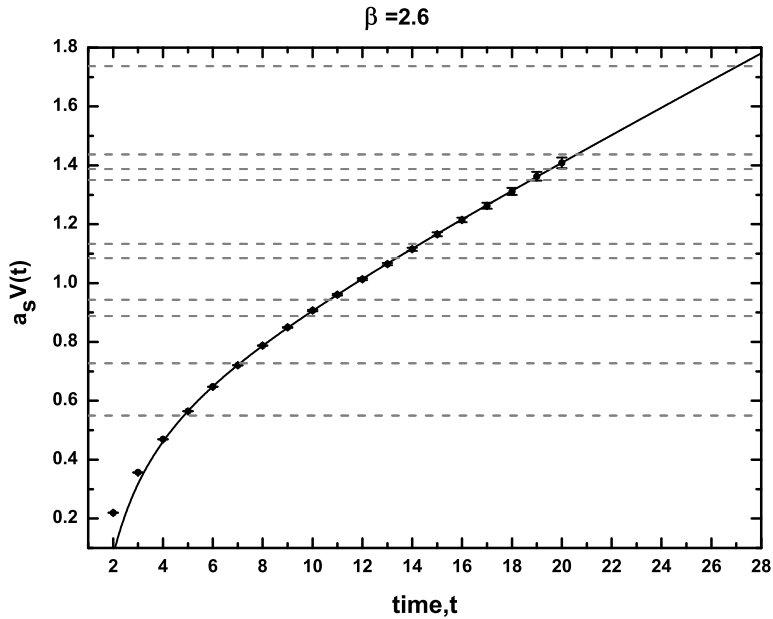}
 \end{center}
 \caption{Determination of physical anisotropy using the
 side-way potential method. This is the case for $\beta=2.6$.}
 \label{fig:vtvs}
 \end{figure}

 To determine the anisotropy $\xi$, we plot
 $a_sV({\hat{t}\over \xi} a_s)$\ at all values of $\hat{t}$ and
 interpolate in the whole range of $\hat{t}$. With this
 interpolation, a curve is obtained which gives the
 values of $a_sV({\hat{t}\over \xi} a_s)$\ versus $\hat{t}$
 continuously within the range.
 In the same plot, we draw horizontal lines with height of
 $a_sV(\hat{R}' a_s)$ at various values of $\hat{R}'$.
 The crossing of these horizontal lines with the interpolation
 curve then offer the value of $\hat{t}$ such that
 Eq.~(\ref{eq:VT}) is satisfied. This in turn gives an
 estimate for the value of $\xi$.
 This situation is demonstrated in Fig.~\ref{fig:vtvs}.
 The data points are the results of
 $a_sV({\hat{t}\over \xi} a_s)$\ at different values of $\hat{t}$.
 The interpolation curve is also shown. Since our temporal lattice
 spacing is finer, usually linear interpolation is adequate.
 The horizontal dashed lines correspond to the
 values of $a_sV(\hat{R}' a_s)$ at various values of $\hat{R}'$.
 The crossing points then finally give the results for
 the renormalized anisotropy.

 The anisotropy obtained in this way depend weakly on
 the values of $\hat{R}'$ chosen to make the connection.
 We choose to quote values of $\xi$ obtained at $\hat{R'}=2$
 as our final estimates since they are most stable and have
 less errors. Finally, we can determine the physical anisotropy
 $\xi$ for all values of $\beta$ that have been simulated.
 \begin{figure}[htb]
 \begin{center}
 \includegraphics[width=12.0cm,angle=0]{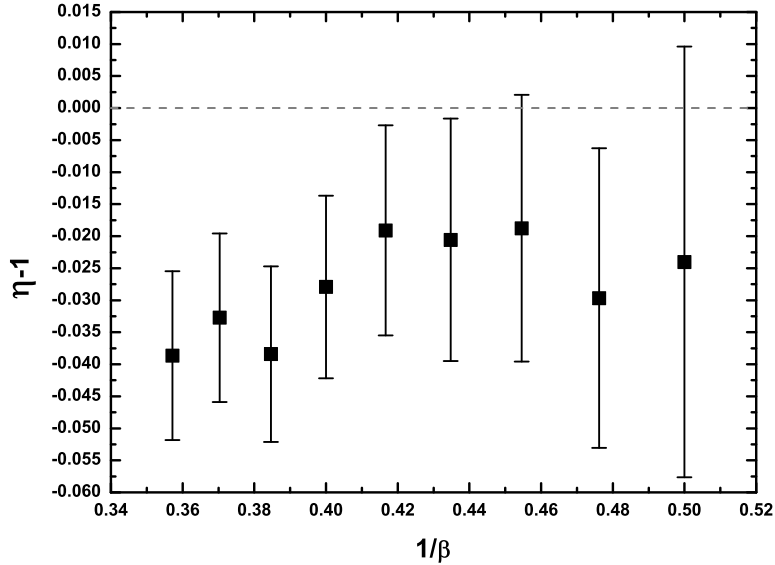}
 \end{center}
 \caption{The effects of renormalization for the anisotropy is shown.
 We plot $\eta=\xi/\xi_0$ at various values of $\beta$, ranging
 between $1.9$ and $2.8$.}
 \label{fig:eta}
 \end{figure}
 In Fig.~\ref{fig:eta}, we plot the ratio of the
 renormalized anisotropy to the bare anisotropy: $\eta=\xi/\xi_0$
 versus $\beta$. It is seen that the relative renormalization
 of the anisotropy parameter is very mild, less than a few percent,
 within the whole range
 of $\beta$. This is mainly due to tadpole improvement. It is
 known that at intermediate values of bare gauge coupling,
 the renormalization can be as large as $30$\% without
 tadpole improvement.

 It is also interesting to compare our measured values with
 the result from perturbation theory. According to
 Ref.~\cite{drummond02:aniso_xi,drummond03:aniso_xi}, the renormalized anisotropy
 can be well fitted with the formula:
 \be
 \eta=\left({u_s\over u_t}\right)\left[1+\left(0.0955-{0.0702\over\chi_0}
 -{0.0399\over\chi^2_0}\right)g^2_0\right]\;,
 \;,
 \ee
 where the "boosted" bare coupling $g^2_0$ and the
 boosted bare anisotropy $\chi_0$ are given by:
 \be
 \beta={6u^3_su_t\over g^2_0}\;,\;\;
 \chi_0=\xi_0{u_s\over u_t}\;.
 \ee
 In our simulation, we find $u_t\simeq 1$ and when plugged
 into the above formulae, qualitative agreement between our measured
 values and the perturbative results are found. However, our
 measured values  of $\eta$ are closer to unity than the
 predictions from perturbation theory.

\section{Conclusions}

 In this paper, we present a numerical study of the static quark
 anti-quark potential using anisotropic
 lattice gauge action. Using the static quark potential obtained
 from Wilson loops, we determine the physical scale of the
 lattice spacing for a range of $\beta$ values. This will set up
 a direct correspondence between the value of $\beta$ and
 the physical lattice spacing. We also measure the renormalized
 anisotropy of the lattice using the side-way potential
 method. It is found that the renormalization effect for
 the anisotropy is small over the whole range of
 $\beta$ that have been studied, typically below 5\%. This is due
 to the use of tadpole improvement. The results obtained in
 this study will be useful for further applications of
 anisotropic lattices on other issues like the
 hadron-hadron scattering.

 \section*{Acknowledgments}

 We would like to thank Prof. H.~Q.~Zheng,
 Prof. S.~L.~Zhu and Prof. S.H.~Zhu of Peking University for helpful
 discussions.


\end{document}